\documentclass[12pt,aps,letterpaper,preprintnumbers,amsmath,amssymb,nofootinbib,notitlepage]{revtex4-1}

\usepackage{graphicx}   
\usepackage{longtable}
\usepackage{lineno,hyperref}
\usepackage{amsmath, amssymb}

\usepackage{epstopdf} 
\usepackage{xcolor}
\usepackage[normalem]{ulem}
\definecolor{awesome}{rgb}{1.0, 0.13, 0.32}

\begin{document}

	\title{Exactness of SWKB for Shape Invariant Potentials}
	
	\author{Asim Gangopadhyaya}
	\email{agangop@luc.edu}
	\author{Jeffry V. Mallow}
	\email{jmallow@luc.edu}
	\author{Constantin Rasinariu}
	\email{crasinariu@luc.edu}
	\author{Jonathan Bougie}
	\email{jbougie@luc.edu}
	\affiliation {Department of Physics, Loyola University Chicago, Chicago, IL 60660, U.S.A}
	
	\date{\today}
		
\begin{abstract}
	The supersymmetry based semiclassical method (SWKB) is known to produce exact spectra for  conventional shape invariant potentials. In this paper we prove that this exactness follows from their additive shape invariance.
	
	\noindent
	{\bf keywords:}
	Supersymmetric Quantum Mechanics, Shape Invariance, Exactly Solvable Systems, Semiclassical Approximation, SWKB
\end{abstract}

\maketitle

\section{Introduction}

The conditions under which semiclassical approximations such as the WKB method yield exact results for quantum-mechanical systems has long been a topic of interest \cite{Dunham,Langer,Bailey,Froman,Krieger1,Krieger2,Krieger3,Bender}.
In 1985, Comtet et al., in the context of supersymmetric quantum mechanics (SUSYQM), proposed a semiclassical quantization condition \cite{Comtet} 
\begin{equation}
\int_{x_1}^{x_2} \sqrt{E_n-W^2(x,a)}\quad {\rm d}x =  n \pi\hbar~, \quad \mbox{where}~ n=0,1,2,\cdots \label{eq:swkb}~
\end{equation}
that generated exact spectra for several solvable systems. Here $W(x,a)$, the superpotential, is connected to the potential energy given by $V_-(x,a) = W^2 - \hbar/\sqrt{2m} \,\, dW/dx$ and limits ${x_1}$ and ${x_2}$ are  given by $W(x_i,a)=\pm \sqrt{E_n}$.
This  quantization condition is known as the Supersymmetric WKB method (SWKB). In 1986, Dutt et al. showed  that the SWKB condition generated exact spectra for all solvable systems  known at that time \cite{Dutt}. It was later shown that this set comprised all $\hbar$-independent shape invariant superpotentials \cite{Bougie2010, symmetry}. 

Even though SWKB quantization  has been found on a case-by-case basis \cite{Dutt_SUSY} to be exact for all $\hbar$-independent shape invariant potentials, there was no general underlying principle to explain it.
It has been conjectured \cite{Khare1989, Dutt91}, but not proved that shape invariance is the source of this SWKB exactness. In this paper we demonstrate that additive shape invariance is sufficient to prove SWKB exactness for all conventional potentials.

\section{Supersymmetric Quantum Mechanics}
\label{sec:SUSYQM}

In SUSYQM \cite{Witten,Solomonson,CooperFreedman, Cooper-Khare-Sukhatme, Gangopadhyaya-Mallow-Rasinariu},  a hamiltonian is written as a product of two differential operators,   
${\cal A}^\pm = \mp \, \frac\hbar{2m}\frac{d}{dx}+ W(x,a)$ that are hermitian conjugates of each other. Setting $2m=1$, the product ${\cal A}^+\!\cdot {\cal A}^-$ generates the hamiltonian  
\begin{equation}
H_-={\cal A}^+\!\cdot {\cal A}^-
=\left(  -\hbar \frac{d}{dx}+ W(x,a) \right)  
\,\left( \hbar \frac{d}{dx}+ W(x,a)\right)
=  -\hbar^2 \frac{d^2}{dx^2}+ V_-(x,a) ~, 
\label{A+A-}
\end{equation}
where the potential $V_-(x,a)$ is given by 
\begin{equation}
V_-(x,a) = W^2(x,a)  -  \hbar \frac{dW}{dx}.
\end{equation} 
The function $W(x,a)$ is known as the superpotential of the system.  Due to the semi-positive definite nature of the hamiltonian $H_-$, its eigenvalues $E^{-}_{n}$ are either positive or zero. If the lowest eigenvalue $E^{-}_{0}\neq 0$, the system is said to have broken supersymmetry. Several authors suggested a modified version of SWKB quantization for systems with broken supersymmetry \cite{Eckhardt, Inomata1,Inomata2}. We will assume that our system has unbroken supersymmetry; i.e., the lowest eigenvalue is zero. 

The product ${\cal A}^-\!\cdot {\cal A}^+$ generates another
hamiltonian $H_+= -\hbar^2 \frac{d^2}{dx^2}+V_+(x,a)$ with $V_+(x,a) =
W^2(x,a) + \hbar \frac{d\, W}{dx}$. The two hamiltonians are related: ${\cal A}^+\!\cdot H_+ = H_- \!\cdot {\cal A}^+$ and ${\cal A}^-\!\cdot H_-
= H_+\!\cdot{\cal A}^-$.  These intertwinings lead to the following relationships among the eigenvalues and eigenfunctions of these ``partner'' hamiltonians: 
\begin{eqnarray}
E^{-}_{n+1} =E^{+}_{n}, \quad \mbox{where}~ n=0,1,2,\cdots~ \label{eq:eigenvalue}
\end{eqnarray}
\begin{eqnarray}
\frac{~~~{\cal A}^- }{\sqrt{E^{+}_{n} }} ~\psi^{(-)}_{n+1} 
= ~\psi^{(+)}_{n}  ~~{\rm and} ~~
\frac{~~~{\cal A}^+}{\sqrt{E^{+}_{n} }}~\psi^{(+)}_{n} 
= ~  \psi^{(-)}_{n+1}~. \label{isospectrality}
\end{eqnarray}
Thus,  knowledge of the eigenvalues and eigenfunctions of one of the hamiltonians automatically gives us their counterparts for the partner hamiltonian.  We note that the hamiltonians $H_\pm$ remain invariant under the following transformations: $W\rightarrow -W$ and  $x\rightarrow -x$. Later in this paper we make use of this property in order to choose signs for some parameters or functions.

\subsection{Shape Invariance}
If the superpotential $W(x,a_i)$ obeys the  ``shape invariance condition"
\cite{Infeld,Miller,gendenshtein1,gendenshtein2},
\begin{equation}
W^2(x,a_i)  +  \hbar \frac{d\, W(x,a_i)}{dx}+g(a_i) = 
W^2(x,a_{i+1})  -  \hbar \frac{d\, W(x,a_{i+1})}{dx}+g(a_{i+1}),~ 
\label{SIC1}
\end{equation}
the spectra for $H_-$ and $H_+$ can be determined algebraically. The eigenvalues and eigenfunctions of $H_-$ are given by 
\begin{eqnarray}
	E_n^{(-)}(a_0)&=&g(a_n)-g(a_0), \label{eq:En}\\
	\psi^{(-)}_{n}(x,a_0&)=& 
	\frac{{\cal A}^+{(a_0)} 
		~ {\cal A}^+{(a_1)}  \cdots  {\cal A}^+{(a_{n-1})}}
	{\sqrt{E_{n}^{(-)}(a_0)\,E_{n-1}^{(-)}(a_1)\cdots E_{1}^{(-)}(a_{n-1})}}~\psi^{(-)}_0(x,a_n)
	~,
\end{eqnarray}
where $\psi^{(-)}_0(x,a_n) = N e^{-\frac1\hbar\, \int^x W(y,a_n)\,dy}$ is the solution of ${\cal A}^-\psi^{(-)}_0 =0 $; i.e., it is the ground-state wavefunction for the eigenvalue $E^{(-)}_{0}= 0$, and $N$ is the normalization constant. 

In this paper we consider only the case of  additive shape invariance: $a_{i+1}=a_{i}+\hbar$.
We further restrict our analysis to the superpotentials $W(x,a_i)$ that have no explicit $\hbar$-dependence; i.e., the $\hbar$-dependence comes in only through parameters $a_i$.
In Ref. \cite{Bougie2010,symmetry}, the authors
showed that in this case, the shape invariance condition reduces to the following two partial differential equations
\begin{eqnarray}
W \, \frac{\partial W}{\partial a} - \frac{\partial W}{\partial x} + \frac12 \, \frac{d g(a)}{d a}= 0~, \label{PDE1} \\
\frac{\partial^{3}}{\partial a^{2}\partial x} ~W(x,a)= 0~,\label{PDE2}
\end{eqnarray} 
and thus demonstrated that Ref. \cite{Dutt_SUSY} provided the complete list of such superpotentials, which we called ``conventional".  Additional shape invariant superpotentials were later found \cite{Quesne1,Quesne2,Quesne2012a, Quesne2012b,Odake1,Odake2, Tanaka,Odake3, Odake4,Ranjani1}, but those were shown to depend explicitly on $\hbar$.

In this paper, we establish that Eqs. (\ref{PDE1}) and (\ref{PDE2}) lead to the exactness of SWKB for all conventional superpotentials.

\subsection{Three Classes of Conventional Shape Invariant Superpotentials}\label{sec:classes}
To prove SWKB exactness from the shape invariance condition for conventional superpotentials, we begin by classifying these superpotentials based on their mathematical form.
From Eq. (\ref{PDE2}), the general form of all such superpotentials is  \cite{Bougie2010,symmetry}
\begin{equation}
W(x,a)= a\,f_1(x) +f_2(x)+u(a) \label{eq:superpotential} ~.
\end{equation}
This form of a typical conventional superpotential derived from Eq. (\ref{PDE2}) was conjectured by Infeld et al. \cite{Infeld}, and examined by others \cite{Ramos99,Ramos00}.
Note that $f_1(x)$ and $f_2(x)$ cannot both be constant, or $W$ would yield trivial potentials with no $x-$dependence. The following three classes of superpotential comprise all possible forms for $W$. Class I: $f_1=\mu$, a constant; Class II: $f_2=\mu$, a constant;  Class III: $f_1$ and $f_2$ both have nonzero $x-$dependence. For each class we now determine the properties which follow from additive shape-invariance. 

\subsubsection{Properties of Class I:}

For Class I,  $f_1=\mu$, a constant. In this case, $W(x,a)=\mu \, a + f_2(x)+u(a)$. We can regroup terms by defining $\tilde{u}(a)\equiv u(a)+\mu~a$, so that  $W(x,a)=f_2(x)+\tilde{u}(a)$. Renaming $\tilde{u}$ back to $u$, eq.(\ref{PDE1}) yields $f_2~\dot{u}-f_2'=-\frac{1}{2} \dot{g} - \dot{u}~u$, where dots denote derivatives with respect to $a$ and primes denote derivatives with respect to $x$. The RHS is independent of $x$. Since $f_2$ does not depend on $a$ and cannot be a constant, each side of the equation is equal to a constant, which we call $\epsilon$. Consequently $\dot{u}=\alpha$, so $u=\alpha~a+\beta$ for  constants $\alpha$ and $\beta$.  Finally, we can regroup terms one more time, such that $\tilde{f_2}(x)\equiv f_2(x)+\beta$ and $\tilde{\epsilon}\equiv \epsilon + \alpha\beta,$ and then rename $\tilde{f_2}$ back to $f_2$ and $\tilde \epsilon$ back to $\epsilon$. 

Therefore, Class I superpotentials can be written as $W(x,a)=f_2(x)+\alpha~a$, where $\alpha f_2-f_2'=\epsilon$ for constants $\alpha$ and $\epsilon$.

\subsubsection{Properties of Class II:}

For Class II, $f_2=\mu$, a constant. In this case, $W(x,a)=a~f_1(x) + \mu +u(a)$. We regroup terms such that $u(a)+\mu\rightarrow u(a)$, so that  $W(x,a)=a~f_1(x)+u(a)$. Then Eq.(\ref{PDE1}) requires $a\left(f_1^2-f_1'\right)+f_1\left(u+a\dot{u}\right)=-\frac{1}{2}\dot{g}-\dot{u}~u$. Since $f_1$ is not constant and the right hand side is $x$-independent, this requires $u+a\dot{u}=a~\alpha$, so $u=\alpha a/2+B/a$ for constants $B$ and $\alpha$. Similar to Class I, we can shift $f_1$ by the constant $\alpha/2$ to absorb the $\alpha a/2$ term into $a f_1$. 

Therefore, Class II superpotentials can be written as $W(x,a)=a f_1(x)+B/a$, where $f_1^2-f_1'=\lambda$ for constants $B$ and $\lambda$.

\subsubsection{Properties of Class III:}

For Class III, neither $f_1$ nor $f_2$ is constant. We first note that if $f_2$ is of the form $f_2(x)=\nu f_1(x)+\mu$ for any constants $\mu$ and $\nu$, then with a redefinition of $a\to a+\nu$, $W$ can be considered equivalent to a Class II superpotential in which $f_2$ is a constant. 
Similarly, if $u(a)$ is linear in $a$, this is equivalent to the case $u(a)=0$ by regrouping terms. 

With these assumptions, we substitute the form of $W$ from Eq.(\ref{eq:superpotential}) in Eq. (\ref{PDE1}) and get
\begin{eqnarray}
\label{eq:fourterms}
	-a \left( f_1^2-f_1^\prime\right) + 
	\left( f_2^\prime-f_1f_2\right) 
	- f_1\left(u+a \dot{u}\right) - \dot{u}f_2 = g/2+\dot{u} u~.
\end{eqnarray}
The first two terms in this equation are respectively linear in $a$ and independent of $a$. Therefore, if there is any nonlinearity in $a$ on the right hand side of the equation, it could only come from the third and fourth terms of the left hand side. However, the right hand side of this equation is independent of $x$; since $f_1$ and $f_2$ are not constant and are linearly independent, the $x$-dependence in the third term cannot be canceled by the fourth term, and vice versa. Consequently, the coefficients of $f_1$ and $f_2$ in terms three and four must each be linear functions of $a$.

Linearity in $a$ of the $\dot{u}f_2$ term in Eq. (\ref{eq:fourterms}) implies $u=\mu a^2/2 + \nu a + \gamma$. Then $u+ a \dot{u}=3\mu a^2/2 + 2\nu a + \gamma$, which is linear in $a$ only if $\mu=0$. Thus $u$ itself is linear in $a$, so we can set $u=0$. 
We then have $-a \left( f_1^2-f_1^\prime\right) +  \left( f_2^\prime-f_1f_2\right) = g/2+\dot{u} u$. Since the right-hand-side is independent of $x$, this requires $f_1^2-f_1^\prime =\lambda$ and $f_1 f_2-f_2^\prime =\varepsilon$ for constants $\lambda$ and $\varepsilon$.

Therefore, Class III superpotentials can be written as $W(x,a)=a f_1(x) + f_2(x)$, where $f_1^2-f_1^\prime =\lambda$ and $f_1 f_2-f_2^\prime =\varepsilon$ for constants $\lambda$ and $\varepsilon$.

\section{Exactness of SWKB}
\label{sec:Exact}
In this section we will show that for the three classes defined above, the definite integral of Eq. (\ref{eq:swkb}) is $n \pi\hbar$. Let us define a function $I(a,n,\hbar)$ by
\begin{equation}
I(a,n,\hbar) \equiv \int_{x_1}^{x_2} \sqrt{E_n-W^2(x,a)}\quad {\rm d}x  \label{eq:swkb1}~.
\end{equation}
Since $W(x,a)$ does not depend on $\hbar$, the energy $E_n=g(a+n\hbar)-g(a)$ is the sole source of $n$ and $\hbar$
dependence of the integrand ${\cal F}(x)= \sqrt{E_n-W^2(x,a)}$. We will prove that $I(a,n,\hbar)=n\pi\hbar$.

First, we note that for $n=0$, $E_0=0$. Hence, ${x_1}$ and ${x_2}$, the roots of  $W(x)=\pm \sqrt{E_n}$, are equal, so $I(a,0,\hbar) =0$. Thus Eq. (\ref{eq:swkb}) holds for $n=0$.

Second, we observe that for all finite values of $a$ and $n$, $\lim_{\hbar \rightarrow 0} E_n \rightarrow 0$. Thus for all $n$, $I(a,n,0) =0$, so if we Taylor expand $I(a,n,\hbar)$ in powers of $\hbar$, there would be no $\hbar$-independent term. I.e., 
	\begin{eqnarray}
	I(a,n,\hbar) = \sum_{k=1}^{\infty} c_k(a,n) \, \hbar^k~.
	\end{eqnarray} 
We now compute the first derivative $\frac{\partial I(a,n,\hbar)}{\partial \hbar}$ for any value of $n$:
	\begin{equation}
	\frac{\partial I(a,n,\hbar)}{\partial \hbar} 
	= \frac{\partial x_2}{\partial \hbar} {\cal F}(x_2) -
	\frac{\partial x_1}{\partial \hbar} {\cal F}(x_1) + 
	\frac{1}{2}\frac{\partial E_n}{\partial \hbar} 
	\int_{x_1}^{x_2} \frac{{\rm d}x}{\sqrt{E_n-W^2(x,a)}\quad}\label{eq:swkb2a}.
\end{equation}
Integrating it, we will determine $I(a,n,\hbar)$. Since ${\cal F}(x)$ vanishes at points ${x_1}$ and ${x_2}$,
\begin{equation}
	\frac{\partial I(a,n,\hbar)}{\partial \hbar} = \frac{1}{2}\frac{\partial E_n}{\partial \hbar} 
	\int_{x_1}^{x_2} \frac{{\rm d}x}{\sqrt{E_n-W^2(x,a)}\quad}~, \label{eq:swkb2}
	\end{equation} 
which is the starting point for proof of SWKB exactness for conventional superpotentials.

Note that if $W(x,a)$ were to have an intrinsic dependence on $\hbar$, as is the case for the extended superpotentials \cite{Quesne1,Quesne2,Quesne2012a, Quesne2012b,Odake1,Odake2, Tanaka,Odake3, Odake4}, Eq. (\ref{PDE2}) would not hold and $W$ would not be restricted to the three classes above, which subsume all the conventional superpotentials. In this case we would have an extra term in Eq. (\ref{eq:swkb2a}). Thus, $\hbar$-dependence of $W$ could impact the exactness of SWKB. For example, a numerical computation \cite{Bougie2017} showed that SWKB was not exact for the extension of the 3D-Oscillator \cite{Quesne1}.  

Next, we will prove that $\partial I(a,n,\hbar)/\partial \hbar=n\pi$, hence $I(a,n,\hbar)=n\pi\hbar$ for the three classes enumerated in Sec.\ref{sec:classes}.

\subsection{Class I}
For this class, we found in Sec.\ref{sec:classes} that $W(x,a)=f_2(x)+\alpha~a$, where $\alpha f_2-f_2'=\epsilon$ for constants $\alpha$ and $\epsilon$. We consider two cases in this class, $\alpha= 0$ and $\alpha\neq 0$.

\subsubsection{Class IA: $\alpha=0$}

In this case, $W(x,a)=f_2(x)$, where $W'(x)=f_2'(x)=-\varepsilon$. From Eq.~\ref{PDE1} $ dg/da = -2 \varepsilon$. To avoid level crossing, $dg/da$ must be positive. We define $\omega\equiv-2\varepsilon>0$, so $dg/da=\omega$,   $E_n=n\omega \hbar$,  therefore $\partial E_n/\partial \hbar=n\omega$. We now solve Eq.~\ref{eq:swkb2} with these values:
\begin{equation}
\frac{\partial I(a,n,\hbar)}{\partial \hbar} =\frac{n\omega}{2}	\int_{x_1}^{x_2} \frac{{\rm d}x}{\sqrt{E_n-W^2(x)}}\nonumber,\end{equation} 
where $x_1$ and $x_2$ are given by solutions to $E_n-W^2=0$. We change the integration variable to obtain
\begin{equation}
\frac{\partial I(a,n,\hbar)}{\partial \hbar} =\frac{n\omega}{2}	\int_{-\sqrt{E_n}}^{\sqrt{E_n}} \frac{2d W}{\omega\sqrt{E_n-W^2(x)}}=n\int_{-\sqrt{E_n}}^{\sqrt{E_n}} \frac{d W}{\sqrt{E_n-W^2(x)}}=n\pi\nonumber.\end{equation}

\subsubsection{Class IB: $\alpha\neq 0$}

\begin{figure}[htb]
	\centering
	\includegraphics[width=0.9\linewidth]{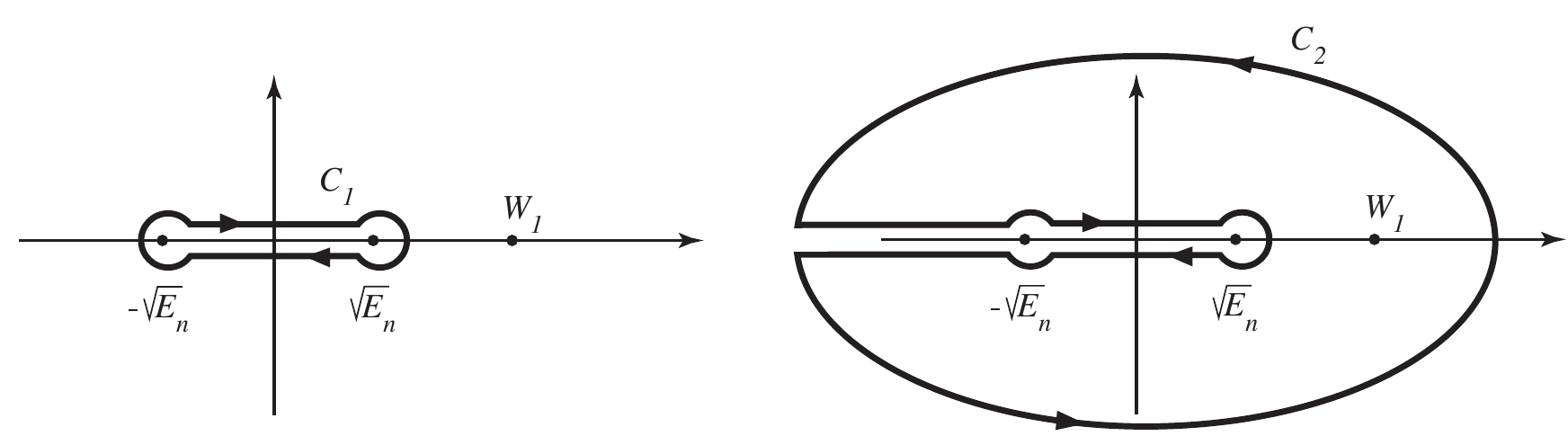}
	\caption{The contour includes one pole at $W_1=\alpha \, a$}
	\label{fig:contour4integration-classi-asim}
\end{figure}

In this case, $W(x,a)=f_2(x)+\alpha\,a$, where $\alpha f_2-f_2'=\epsilon$ for nonzero $\alpha$. With a redefinition $a+\varepsilon/\alpha \rightarrow  a$, we can set  $\varepsilon=0$ and $f_2-\epsilon/\alpha\rightarrow f_2$. Thus, we have 
\begin{eqnarray}
W(x,a) = \alpha\,a + f_2(x) \quad\mbox{and} \quad \frac{\partial W}{\partial x}=f_2'(x)=\alpha f_2(x)=
 \alpha\, (W-\alpha\,a)~.
\end{eqnarray} 
 Since $f_2' = \alpha f_2$, $f_2$ cannot be zero at any point, or it would be zero everywhere. Hence $W'$ must have a definite sign, which for unbroken supersymmetry must be positive. This implies that $\alpha f_2>0$, so $\alpha$ and $f_2$ must have the same sign. Without loss of generality
 \footnote{Here we have used the fact that the symmetry operations $W\rightarrow -W$ and  $x\rightarrow -x$ discussed in Sec. \ref{sec:SUSYQM}, do not change the  value of the integral of Eq. (\ref{eq:swkb1}). }, 
 we assume $\alpha<0$; consequently, $f_2<0$.  From Eq. (\ref{PDE1}), we have $dg/da = -2 \alpha^2\,a$.  Because  $dg/da>0$  we must have $a<0$ (and $a+n\hbar<0$ for all bound states), thus  $W < \alpha\,a$.  By integrating $g$, we get $E_n = g(a+n\hbar)-g(a) = \alpha^2\,a^2 -  \alpha^2 (a+n\hbar)^2$. Then
\begin{eqnarray}
\frac{\partial I(a,n,\hbar)}{\partial \hbar} = \frac12 
\frac{\partial E_n}{\partial \hbar} 
\int_{x_1}^{x_2} \frac{{\rm d}x}{\sqrt{E_n-W^2(x,a)}}
=
\frac1{2\alpha }  \frac{\partial E_n}{\partial \hbar} 
\int_{-\sqrt{E_n}}^{\sqrt{E_n}}\frac{{\rm d}W}{(W-\alpha\,a)\sqrt{E_n-W^2(x,a)}}. \nonumber		
\end{eqnarray} 
We carry out the integration in the complex $W$ plane, as shown in Fig. \ref{fig:contour4integration-classi-asim}. 

The contour includes a pole at $W_1=\alpha \, a$, and a cut from ${-\sqrt{E_n}}$ to ${\sqrt{E_n}}$.  The partial derivative $\frac{\partial I(a,n,\hbar)}{\partial \hbar} $ is then given by
\begin{eqnarray} 
\frac{\partial I(a,n,\hbar)}{\partial \hbar} 	&=&
\frac1{4\alpha }  \frac{\partial E_n}{\partial \hbar} 
\oint\frac{{\rm d}W}{(W-\alpha \, a)\sqrt{E_n-W^2(x,a)}} 
\nonumber\\
\nonumber\\
&=&
\frac1{4\alpha }  \frac{\partial E_n}{\partial \hbar} 
\frac{2\pi \, i}{\sqrt{E_n-\alpha^2\,a^2 }} =
\frac1{4\alpha }  \frac{\partial E_n}{\partial \hbar} 
\frac{2\pi \, i}{\sqrt{- \alpha^2(a+n\hbar)^2}}   
\nonumber\\
\nonumber\\
&=&
\frac{n\pi}{2\alpha^2}  \,  
\frac{-2\alpha^2(a+n\hbar)}{-(a+n\hbar) } =
n\pi~,
\end{eqnarray} 
where we substituted ${\partial E_n}/{\partial \hbar}  = {-2\alpha^2(a+n\hbar)n} $.

\subsection{Class II}

From Sec.\ref{sec:classes}, this class is of the form $W(x,a)=a f_1(x)+B/a$, where $f_1^2-f_1'=\lambda$ for constants $B$ and $\lambda$.
From \ref{PDE1}, this requires
\begin{equation}
\label{eq:f2const}
\frac{dg}{da} = \frac{2B^2}{a^3} - 2 \lambda \, a;~~ g(a)=-\frac{B^2}{a^2} - \lambda\, a^2 ~,
\end{equation}
and thus
$
E_n=\frac{B^2}{a^2} - \frac{B^2}{(a+n\hbar)^2} +\lambda\left[\, a^2-(a+n\hbar)^2\right]
$. From $dg/da = \frac{2B^2}{a^3} - 2 \lambda \, a$, we see that if $\lambda\leq 0$, we must have $a>0$. For $\lambda>0$, we have two cases:  $a>0$ and ${B^2} >  \lambda \, a^4$, or  $a<0$ and ${B^2} <  \lambda \, a^4$. 

Using $W' = a(f_1^2 -\lambda) $, we change the integration variable:
$$
dx = \frac{a\, dW}{ \left( W-B/a\right)^2 -\lambda\, a^2  }~.
$$
Then Eq. (\ref{eq:swkb2}) becomes
\begin{equation}
\label{eq:dIdh}
\frac{\partial I(a,n,\hbar)}{\partial \hbar} = 
\frac{a}{2}\frac{\partial E_n}{\partial \hbar} \int_{-\sqrt{E_n}}^{\sqrt{E_n}}
\frac{dW}{\left[\left( W-B/a\right)^2 -\lambda\, a^2 \right]
\sqrt{E_n - W^2}}~.
\end{equation}
Here we have two cases: $\lambda= 0$ and $\lambda\neq 0$. 

\subsubsection{Class IIA: $\lambda=0$}
For $\lambda=0$, since $f_1' = f_1^2$, we see that if $f_1=0$, at one point, it must be zero at all points.  Hence $f_1$ must have a definite sign everywhere. Without loss of generality, we choose $f_1<0$. Then, since $W = af_1(x) + B/a$ must change sign to preserve supersymmetry, we must have $B>0$.  Hence,  Eq. (\ref{eq:dIdh}) becomes
\begin{equation}
	\frac{\partial I(a,n,\hbar)}{\partial \hbar} = \frac{anB^2}{(a+n\hbar)^3}
	\int_{-\sqrt{E_n}}^{\sqrt{E_n}}
	\frac{dW}{\left( W-B/a\right)^2 \sqrt{E_n - W^2}}~~,
\end{equation}
which integrates to 
\begin{equation} 
	\frac{\partial I(a,n,\hbar)}{\partial \hbar} 
	= \frac{anB^2}{(a+n\hbar)^3}\frac{a^2 B \pi}{(B^2-a^2 E_n)^{3/2}} =n\pi~,
\end{equation}
where we used $E_n=\frac{B^2}{a^2} - \frac{B^2}{(a+n\hbar)^2} $.

\subsubsection{Class IIB: $\lambda\ne0$}
This class splits into two cases: $\lambda >0 $ and $\lambda <0 $. 

We will consider first the case when $\lambda >0 $. Because 
\begin{equation}
\label{eq:en-l>0}
	E_n=\frac{B^2}{a^2} - \frac{B^2}{(a+n\hbar)^2} +\lambda \left[\, a^2-(a+n\hbar)^2\right]~,
\end{equation}
Eq. (\ref{eq:dIdh}) becomes
\begin{equation}
\label{eq:dIdh1234}
\frac{\partial I(a,n,\hbar)}{\partial \hbar} = 
\frac{a}{2}
\left[\frac{2 B^2 n}{(a + n \hbar )^3}-2 n \lambda (a+n\hbar)\right]
\int_{-\sqrt{E_n}}^{\sqrt{E_n}}
\frac{dW}{(W-W_1)(W-W_2)
	\sqrt{E_n - W^2}}
\end{equation}
where the simple poles $W_1= B/a+a\sqrt{\lambda}$ and $W_2=B/a-a\sqrt{\lambda}$ are both greater than $\sqrt{E_n}$~, and
$B>\sqrt{\lambda } (a+ n \hbar)^2$.
To compute this integral we first observe that it can be written as a sum of two integrals  which can be evaluated independently. Using Eq. (\ref{eq:en-l>0}), we obtain\footnote{The computation shown assumes $a>0$. The case $a<0$ yields the same result.} 
\begin{eqnarray*}
J &&\equiv \int_{-\sqrt{E_n}}^{\sqrt{E_n}} \frac{dW}{(W-W_1)(W-W_2)\sqrt{E_n - W^2}}   \\
&&=\frac{1}{W_1-W_2}\int_{-\sqrt{E_n}}^{\sqrt{E_n}} \frac{dW}{(W-W_1)\sqrt{E_n - W^2}} 
+  \frac{1}{W_2-W_1}\int_{-\sqrt{E_n}}^{\sqrt{E_n}} \frac{dW}{(W-W_2)\sqrt{E_n - W^2}} \\
&&=-\frac{\pi  (a+n\hbar)}{2 a \sqrt{\lambda}\, \left(B+ \sqrt{\lambda } (a+n\hbar)^2\right) }
+ \frac{\pi  (a+n\hbar)}{2 a \sqrt{\lambda} \left( B-\sqrt{\lambda}\,(a+n\hbar)^2 \right) }
 = ~ \frac{\pi  (a+n\hbar)^3}{a B^2-a \lambda  (a+n\hbar)^4}~.
\end{eqnarray*}
Substituting $J$ back into Eq. (\ref{eq:dIdh1234}) we arrive at
\begin{equation}
\frac{\partial I(a,n,\hbar)}{\partial \hbar} = n\pi~.
\end{equation}

Let us now consider the case when $\lambda =-\mu <0 $. Then Eq. (\ref{eq:dIdh}) becomes
\begin{equation}
\label{eq:dIdh2}
\frac{\partial I(a,n,\hbar)}{\partial \hbar} = 
\frac{a}{2}
\left[\frac{2 B^2 n}{(a + n \hbar n)^3}+2 n \mu (a+n\hbar)\right]
\int_{-\sqrt{E_n}}^{\sqrt{E_n}}
\frac{dW}{(W-U_1)(W-U_2)
	\sqrt{E_n - W^2}}~,
\end{equation}
where the simple poles are $U_1= B/a +i a \sqrt{\mu}$ and $U_2=B/a-i a\sqrt{\mu}$. Note that this is a real positive integral 
\footnote{ The case $\lambda<0$ requires $W'>0$ at every point, so the derivative  $(W^2)' = 2 W W' = 0$ at only those points where $W=0$, and this happens only at one point $x_0$. At $x_0$ the second derivative is positive, hence $W^2$ has only one extremum, a minimum at $x_0$. This implies that the integral $\int_{x_1}^{x_2}\left[E_n-W^2(x,a)\right]^{-1/2}dx$ is real and positive, as the integrand is real and positive at every point in the domain. }. 
The complex factorization in Eq. (\ref{eq:dIdh2}) was done in order to carry out the calculations in the complex $W$ plane, as illustrated in Fig. (\ref{fig:C1C2a}). We obtain for the integral in Eq. (\ref{eq:dIdh2}):
\begin{figure}[htb]
	\centering
	\includegraphics[width=\linewidth]{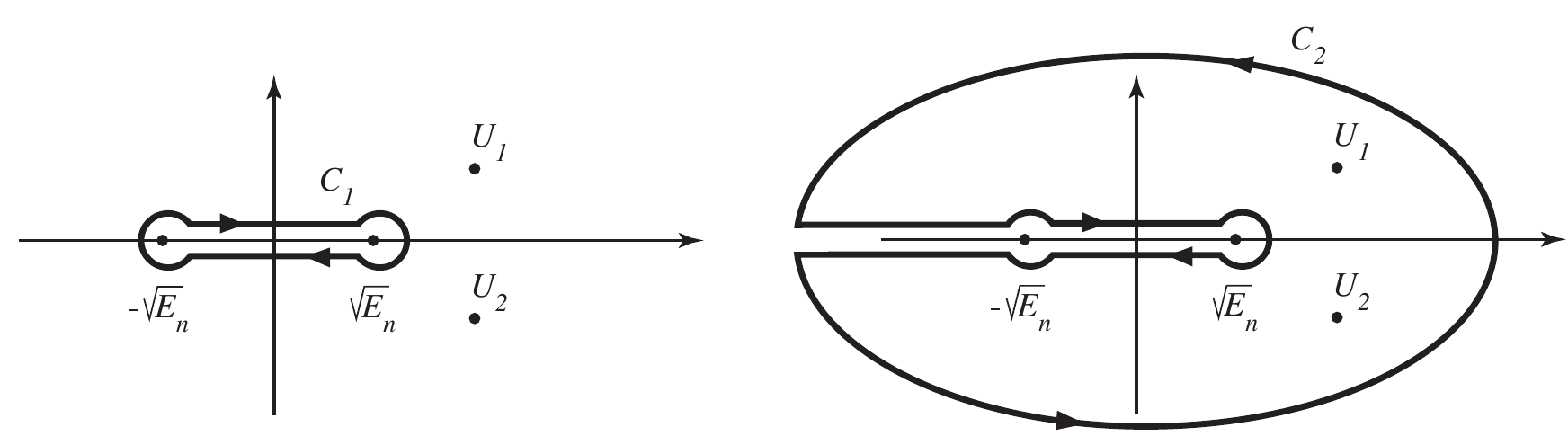}
	\caption{Complex integration for $\lambda=-\mu < 0$ case. The contour $C_2$ includes simple poles
		$U_1= B/a+ i a\sqrt{\mu}$ and $U_2=B/a- i a\sqrt{\mu}$.}
	\label{fig:C1C2a}
\end{figure}
\begin{eqnarray}
\label{eq:L.lt.0}
J \equiv && \int_{-\sqrt{E_n}}^{\sqrt{E_n}}
\frac{dW}{(W-U_1)(W-U_2)
	\sqrt{E_n - W^2}} \nonumber \\   
=&&\frac{\pi (a+n\hbar)}{2 a \sqrt{\mu }} \left(
\frac{1}{\sqrt{\left(\sqrt{\mu } (a+n\hbar)^2-i B\right)^2}}
-\frac{1}{\sqrt{\left(\sqrt{\mu } (a+n\hbar)^2+i B\right)^2}}
\right) \nonumber\\
=&&\frac{\pi (a+n\hbar)}{2 a \sqrt{\mu }} \left(
\frac{e_1}{{\sqrt{\mu } (a+n\hbar)^2-i B}}
-\frac{e_2}{{\sqrt{\mu } (a+n\hbar)^2+i B}}
\right)~,
\end{eqnarray}
where $e_1, e_2 = \pm 1$. When substituted into Eq. (\ref{eq:dIdh2}), this gives
\begin{equation}
\frac{\partial I(a,n,\hbar)}{\partial \hbar} = 
\frac{1}{2} \pi  n \left(e_1 - e_2 + \frac{i B (e_1+ e_2)}{\sqrt{\mu } (a+ n\hbar )^2 }  \right)~.
\end{equation}
Since this is a real positive integral, we must have $e_1+e_2 = 0$ and $e_1 = 1$. We arrive at
\begin{equation}
\label{eq:dIdh-l.lt.0}
\frac{\partial I(a,n,\hbar)}{\partial \hbar} = n\pi~.
\end{equation}
 
\subsection{Class III}

For Class III, $W(x,a)=a f_1(x) + f_2(x)$, where $f_1^2-f_1^\prime =\lambda$ and $f_1 f_2-f_2^\prime =\varepsilon$, for constants $\lambda$ and $\varepsilon$. We have two cases in this class, $\lambda= 0$ and $\lambda\neq 0$. We now examine each case separately.

\subsubsection{Class IIIa: $\lambda =0$}

In this case, $f_1'=f_1^2$, hence $f_1$ cannot be zero at any point. Also, $f_2'=f_1\,f_2 -\varepsilon$. The homogeneous equation for $f_2'=f_1\,f_2 -\varepsilon$ is solved by $f_2 = \alpha f_1$. A particular solution is $f_2 = \frac12 \varepsilon/f_1$. Thus, the superpotential takes the form $W=af_1+  \alpha f_1+\frac12 \varepsilon/f_1 = (a+\alpha) f_1+\frac12 \varepsilon/f_1 \equiv af_1+ \frac12 \varepsilon/f_1$, where we have redefined the parameter $a$.  From Eq.~\ref{PDE1}, $dg/da=-2\varepsilon>0$, implies $\varepsilon<0$, which requires that $W'=af_1^2-\varepsilon/2>0$, because $a>0$\footnote{Unbroken supersymmetry requires that $W=0$ for some value of $f_1$, which occurs when $f_1^2=|\varepsilon|/2a$, so  $a>0$.}. So
\begin{eqnarray}
\frac{\partial I(a,n,\hbar)}{\partial \hbar} &=&\frac{1}{2}\frac{\partial E_n}{\partial \hbar}\int_{x_1}^{x_2}\frac{dx}
{
	\sqrt{
		E_n-\left(af_1+ \frac12 \varepsilon/f_1
		\right)^2
	}
}~.
\nonumber
\end{eqnarray}
Changing the integration variable to $f_1$,
\begin{eqnarray}
\frac{\partial I(a,n,\hbar)}{\partial \hbar} &=&\frac{1}{2}\frac{\partial E_n}{\partial \hbar}\int_{f_L}^{f_R}\frac{df_1}
{f_1^2
	\sqrt{
		E_n-\left(af_1+ \frac12 \varepsilon/f_1
		\right)^2
	}
},
\nonumber
\end{eqnarray}
where ${f_L}$ and ${f_R}$ are the turning points on the $f_1$ axis, where the square root in the denominator is zero\footnote{Since $df_1=f_1^2\, dx$, the relative positioning of ${f_L}$ and ${f_R}$ remains the same as $x_1$ and $x_2$.}. Moving to the complex $f_1$-plane,
\begin{eqnarray}
\frac{\partial I(a,n,\hbar)}{\partial \hbar} 
&=&\frac12 \frac{1}{2}\frac{\partial E_n}{\partial \hbar}\oint \frac{df_1}
{f_1^2
	\sqrt{
		E_n-\left(af_1+ \frac12 \varepsilon/f_1
		\right)^2
	}
}
\nonumber\\\nonumber\\
&=&\frac14 {\frac12\, (-2\varepsilon)\, }  n \oint \frac{df_1^2}{f_1^2
	\sqrt{
		E_nf_1^2-\left(af_1^2+ \frac12 \varepsilon\right)^2
	}
}=-\frac14 \varepsilon n (2\pi i) \frac{2}{-i\, \epsilon} = n\pi\nonumber.
\end{eqnarray}

\subsubsection{Class IIIb: $\lambda \neq 0$}

In this case, $f_1'=f_1^2-\lambda$, $f_2'=f_1\,f_2 -\varepsilon$, and $W=af_1+f_2$. The homogeneous and particular solutions for $f_2$ are $\beta \sqrt{f_1^2-\lambda}$  and $f_{1}\left(\frac{\varepsilon}{\lambda}\right)$ respectively \footnote{ The homogeneous solution is
	$
	f_2 	=\beta \exp{\left[ \int f_1 \, dx\right]}  = \beta \exp{\left[ \int f_1 \, \frac{df_1}{f_1^2-\lambda}\right]}  =\beta \exp{\left[ \frac12  \int \frac{df_1^2}{f_1^2-\lambda}\right]}  = \beta \sqrt{f_1^2-\lambda} ~.
	$
}. Thus,  with a redefinition of the parameter $a$, we get $W=af_1+f_{1}\left(\frac{\varepsilon}{\lambda}\right)+\beta \sqrt{f_1^2-\lambda}=
\left( a+\frac{\varepsilon}{\lambda}\right)\,f_{1}+\beta \sqrt{f_1^2-\lambda} \equiv af_1+\beta \sqrt{f_1^2-\lambda}$.

From Eq.~\ref{PDE1} we have $g(a)=-\lambda\, a^2 $, so $E_n=\lambda\left[a^2-(a+n\hbar)^2\right]$, and $\partial E_n/\partial\hbar=-2n\lambda(a+n\hbar)$. To ensure the order  $E_{n+1}>E_n>E_{n-1}$, we must have $\lambda(a+n\hbar)<0$.

Using the fact that $f_1^2 \neq \lambda$,  we define a function 
$y (x)\equiv \frac{\sqrt{\lambda} -f_1}{\sqrt{\lambda}+f_1}$. Its derivative is given by $\frac{dy}{dx}=2\sqrt{\lambda} y$,  which yields $f_1 = \sqrt{\lambda}\left(\frac{y-1}{y+1} \right) $. We now define two functions ${\cal{S}}(x)\equiv\frac{y^{1/2}-y^{-1/2}}{2\sqrt{\lambda}}$, and ${\cal{C}}(x)\equiv\frac{y^{1/2}+y^{-1/2}}{2}$, which satisfy the identities: 
$$
\begin{array}{lll}
{d{\cal{C}}}/{dx}=\lambda {\cal{S}}~, & {d{\cal{S}}}/{dx}={\cal{C}}~, &
{\cal{C}}^2(y)-\lambda \,{\cal{S}}^2(y) = 1~, \\
2\,{\cal{C}}(y)\,{\cal{S}}(y) = {\cal{S}}(y^2)~, \quad \quad &  {\cal{C}}^2(y)+\lambda \,{\cal{S}}^2(y) = {\cal{C}}(y^2)~.
\end{array}
$$
In terms of these variables, $f_1 = -\lambda \frac{{\cal{S}}}{{\cal{C}}}$ and  $f_2 = \beta \sqrt{f_1^2-\lambda} =  \frac{\beta}{{\cal{C}}}$, where $\beta$ is a constant.  Now, we proceed to compute ${\partial I(a,n,\hbar)}/{\partial \hbar}$ for this case.
\begin{eqnarray}
\frac{\partial I(a,n,\hbar)}{\partial \hbar}&=&\frac{1}{2}\frac{\partial E_n}{\partial \hbar}\int_{x_1}^{x_2}\frac{dx}{\sqrt{E_n-W^2}}~=~ -n\lambda(a+n\hbar)\int_{x_1}^{x_2}\frac{dx}{\sqrt{E_n-(af_1+f_2)^2}}\nonumber~,
\end{eqnarray}
which can be written as 
\begin{eqnarray}
\frac{\partial I(a,n,\hbar)}{\partial \hbar}&=&-n\lambda(a+n\hbar)\int_{x_1}^{x_2}\frac{dx}{\sqrt{E_n-(-\frac{\lambda {\cal{S}} a}{{\cal{C}}}+\frac{\beta}{{\cal{C}}})^2}}\nonumber\\
&=&-\frac12 \, n\lambda(a+n\hbar)\oint \frac{d{\cal{S}}}{\sqrt{E_n\, \left(1+ \lambda{\cal{S}}^2\right) -\lambda^2a^2{\cal{S}}^2+2\lambda {\cal{S}}\beta a-\beta^2}}\nonumber\\
&=&-\frac12 \, n\lambda(a+n\hbar)\oint \frac{d{\cal{S}}}{\sqrt{
		{\cal{S}}^2\left(  \lambda E_n -\lambda^2a^2 \right)  +2\lambda {\cal{S}}\beta a-\beta^2}}\nonumber\\
&=&-\frac12 \, n\lambda(a+n\hbar)\oint \frac{dt}{t\sqrt{
		\left(  \lambda E_n -\lambda^2a^2 \right)  +2\lambda\beta a\, t-\beta^2t^2}}, \quad \quad \mbox{where } t \equiv 1/{\cal{S}}\,; \nonumber\\
&=&-\frac12 \, n\lambda(a+n\hbar)\,\frac{ 2\pi i}{\sqrt{
		-\lambda^2 (a+n\hbar)^2}}\, = -\frac12 \, n\,\frac{ 2\pi(a+n\hbar) }
{\left| a+n\hbar\right|}\, = n\pi~,
\end{eqnarray}
where we have used the constraint $\lambda(a+n\hbar)<0$. 

\section{Conclusion}
In this paper we have proved that the exactness of SWKB for conventional superpotentials follows from the additive shape invariance condition.

\end{document}